\documentclass[rnote]{aa} % for the research notes
\usepackage{graphicx}
\usepackage{txfonts}
\begin{document} 

\title{Polarisation Observations of H$_{2}$O $\mathbf{J_{K_{-1}K_{1}}=5_{32}-4_{41}}$
620.701 GHz Maser Emission with HERSCHEL%
\thanks{\emph{Herschel }is an ESA space observatory with science instruments
provided by European-led Principal Investigator consortia and with
significant participation from NASA. %
}/HIFI in Orion KL}

\author{S. C. Jones\inst{1}, M. Houde\inst{1,2},
M. Harwit\inst{3}, M. Kidger\inst{4}, A. Kraus\inst{5},
C. M\textsuperscript{c}Coey\inst{6}, A. Marston\inst{4},
G. Melnick\inst{7}, K. M. Menten\inst{5}, P.
Morris\inst{8}, D. Teyssier\inst{4}, V. Tolls\inst{7}}

%\maketitle
\institute{University of Western Ontario, Department of Physics and
  Astronomy, London, Ontario, N6A 3K7, Canada \email{ sjone7@uwo.ca}
  \and Division of Physics, Mathematics and Astronomy, California
  Institute of Technology, Pasadena, CA 91125 \and Cornell University,
  Center for Radiophysics \& Space Research, 511 H Street, SW,
  Washington, DC 20024-2725, USA  \and Herschel Science Centre, ESAC,
  European Space Agency, 28691 Villanueva de la Ca\~{n}ada, Madrid,
  Spain \and Max-Planck-Institut f\"{u}r Radioastronomie, Auf dem
  H\"{u}gel 69, 53121 Bonn, Germany \and University of Waterloo,
  Department of Physics and Astronomy, Waterloo, Ontario, N2L 3G1,
  Canada \and Harvard-Smithsonian Center for Astrophysics, 60 Garden
  Street, MS 66, Cambridge, MA 02138, USA \and Infrared Processing and
  Analysis Center, California Institute of Technology, Pasadena, California, USA}

\abstract {The high intensities and narrow bandwidths exhibited by
  some astronomical masers make them ideal tools for studying
  star-forming giant molecular clouds. The water maser transition
  $J_{K_{-1}K_{1}}=5_{32}-4_{41}$ at 620.701 GHz can only be observed
  from above Earth's strongly absorbing atmosphere; its emission has
  recently been detected from space.} 
{We sought to further characterize the star-forming environment of
  Orion KL by investigating the linear polarisation of a source
  emitting a narrow 620.701 GHz maser feature with the heterodyne
  spectrometer HIFI on board the \emph{Herschel} Space Observatory.} 
{High-resolution spectral datasets were collected over a thirteen
  month period beginning in 2011 March, to establish not only the
  linear polarisation but also the temporal variability of the
  source.} 
{Within a $3\sigma$ uncertainty, no polarisation
  was detected to an upper limit of approximately 2\%. These results
  are compared with coeval linear polarisation measurements of
  the 22.235 GHz $J_{K_{-1}K_{1}}=6_{16}-5_{23}$ maser line from the
  Effelsberg 100-m radio telescope, typically a much stronger
maser transition. Although strongly polarised emission is observed for
one component of the 22.235 GHz maser at 7.2 km s$^{-1}$, a weaker
component at the same velocity as the 620.701 GHz maser at 11.7 km
s$^{-1}$ is much less polarised.  } 
{}

\titlerunning{Polarisation Observations of a 620.701 GHz Water Maser
  Transition in Orion KL}
\authorrunning{S.C. Jones et al.}
\maketitle

\section{Introduction\label{sec:Introduction}}

The Kleinmann-Low nebula in the Orion Molecular Cloud 1 (OMC-1) is a
high-mass star-forming region,  the nearest such region in our Galaxy
at 418$\pm$6 pc \citep{Kim2008}. Its line-of-sight velocity relative
to the local standard of rest (LSR) is about 8 km s$^{-1}$
\citep{Garay1989}.  Since the discovery of the 22.235 GHz water
maser transition in the Orion Molecular Cloud \citep{Cheung1969}, the
region has frequently been studied also at higher frequency maser transitions. 
Recently, \citet{Neufeld2013} mapped the theoretically predicted
$J_{K_{-1}K_{1}}=5_{32}-4_{41}$ 620.701 GHz transition in the
Kleinmann-Low Nebula (Orion KL) with \emph{Herschel/}HIFI.  Combining
their observations with collocated 22.235 GHz Effelsberg data, they
were able to verify a maser pumping model arising from collisional
excitation  and spontaneous radiative decay
\citep{Neufeld1991}. Hereafter, we will mostly refer to these
frequencies as 22 GHz and 621 GHz.

The first astronomical detection of the  621 GHz maser
transition of ortho water vapor \citep{Neufeld1991} was made
by \citet{M.Harwit2010}.  It appeared to exhibit a
polarisation of approximately 2\% toward the oxygen-rich supergiant
star VY Canis Majoris, and was a few times less luminous than the
star's ortho-H$_2$O 22 GHz maser. The aforementioned study of
\citet{Neufeld2013} presented the first detection of this transition in
an interstellar region (i.e., Orion KL).
Their maps of this region revealed a spatially localized source
emitting a strong, spectrally narrow emission feature at 621 GHz.
Their discovery of this feature gave rise to the present attempt to
determine whether the 621 GHz feature might be linearly polarised. In
a later search through the \emph{Herschel} archives
\citet{Neufeld2013} also found similarly narrow  621 GHz emission
features emanating from the Orion South condensation and the W49N
region of high-mass star formation.  By then, however, the
\emph{Herschel} mission was ending and a search for signs of
polarisation in these features were no longer possible.  To date, the
Heterodyne Instrument for the Far Infrared (HIFI) \citep{Graauw2010}
onboard the \emph{Herschel} Space Observatory \citep{Pilbratt2010} has
been the sole facility capable of linear polarisation studies of
spectral lines at this frequency. Complementary circular
polarisation studies, however, have remained beyond reach.

Between March 2011 and April 2012, we obtained six sets of
observations of the 621 GHz line along a sight line toward Orion KL
with the  \emph{Herschel/}HIFI instrument.  We complemented these with
observations of the 22 GHz line taken contemporaneously with the
Max-Planck-Institut f\"{u}r Radioastronomie (MPIfR) 100-m telescope in
Effelsberg, Germany. In Section \ref{sec:Observations} we report
the results of these observations. We next present our polarisation
results in Section \ref{sec:Results}, along with a depiction of the temporal
behaviour of the maser line over this year-long period. Finally,
Section \ref{sec:Conclusion} presents our conclusions, while the
tabular summaries of our observations, our maps of the 621 GHz water
maser, and a further discussion on pointing errors can be found in
Appendices \ref{table_summary}, \ref{HIFImaps}, and
\ref{sec:Discussion} (on line material).

\section{Observations\label{sec:Observations}}

\subsection{HIFI Observations\label{sub:HIFI-Observations}}

HIFI enables observations in fourteen frequency bands.  We observed 
the 621 GHz transition in HIFI Band 1B, which covers the range from 562.6 to
628.4 GHz.   Like all of the other HIFI bands, Band 1B houses two channels
sensitive to linearly polarised radiation, respectively, in the horizontal
(H) and vertical (V) directions.  The two channels exhibit peak
sensitivities along directions at angles of $82.5^{\circ}$ (H) and
$-7.5^{\circ}$ (V), relative to the \emph{Herschel}
spacecraft's y-axis, kept close to perpendicular to the ecliptic plane
at all times.  On the sky, the H- and V-beams are offset from each
other by $6\farcs6$, a non-negligible fraction of the $34\farcs4$
full-width-half-magnitude (FWHM) beam
at 620.701 GHz.  Due to potential beam pointing errors, small offsets
in the H- and V-beam from their intended positions may be
expected, leading to uncertainty in the mean beam pointing direction. At the time of writing this was constrained to between $0\farcs8-0\farcs9$ ($1\sigma$). 

Observations were conducted beginning on 14 March 2011 when
a small map was acquired using the HIFI ``on-the-fly mapping'' (OTF)
mode, as part of the HEXOS Guaranteed Time Key Program (PI:
E. Bergin). A total of fifteen Nyquist-sampled pointings
were implemented in a rectangular, 5 by 3 configuration. Spectral data
were obtained with the digital autocorrelation high-resolution spectrometer 
(HRS) and the wide-band spectrometer (WBS). The map was centered at 
(R.A.{[}J2000{]}$=05^{\mbox{h}}35^{\mbox{m}}13^{\mbox{s}}.16$,
decl.{[}J2000{]}$=-05^{\circ}22\arcmin00\farcs5$), and spanned $\simeq21\farcs3$
in Right Ascension and $\simeq53\arcsec$ in Declination. Individual
pointings were separated by $\simeq10\farcs5$ in R.A. and 
$\simeq13\arcsec$ in Declination.  The spectral resolution in the HRS mode
was 0.125 MHz, or, equivalently, a line-of-sight velocity
resolution of 0.06 km s$^{-1}$. The WBS resolution was $\approx1$
MHz or $\approx0.5$ km s$^{-1}$. All observations used an
OFF-source reference position located 14$\arcmin$ from the maser
location.

Following the mapping of 14 March 2011, two pointed observations were
obtained that year, respectively, on 26 March and 8 April. Over this
period the source rotated by $\approx14^{\circ}$ about the spacecraft
line-of-sight.  Observations obtained at different times
were necessary since a minimum of two sets of measurements at separate
source rotation angles are required for polarisation analyses
\citep{M.Harwit2010}.  For pointed observations the central point on
the line separating the H- and V-beam centers was  directed at two
successive positions on the sky, in position switching mode, in order
to place the center of the H polarisation beam for a given integration
to coincide with the center of the V polarisation beam for the
subsequent integration -- thus compensating for the misalignment
between the beams.  

After a preliminary analysis, follow-up observations were executed
on 25 February and 14 April 2012. The first of these, performed
in position switching mode, were centered at the same position cited
above.  Thereafter, on 14 April, the last of the pointed measurements
was obtained, followed on the same day by a small map of the same
dimensions as that of 14 March 2011. While the 2011 observations
lasted 888 s, integrations in 2012 were extended to 1683 s in order to
reduce the overall noise.  The entirety of the investigation is summarised
online in Table \ref{obssum}.

\subsection{Effelsberg Observations}

Ground-based observations were undertaken with the 100-m Effelsberg
radio telescope in tandem with the HIFI observations of 2011 and
2012. Centered at 22 GHz, the $J_{K_{-1}K_{1}}=6_{16}-5_{23}$
transition of ortho water vapour, our observations consisted of linear
polarisation signals obtained from two orthogonal channels of the
$K-$band (1.3 cm) receiver located at the primary focus of the
telescope. The Effelsberg beam profile can be approximated
as a Gaussian with a FWHM of $41\farcs0$. The frequency resolution of each
dataset was 6.104 kHz, corresponding to a velocity resolution of 0.082
km s$^{-1}$, and the spectra were calibrated using corrections for
gain-elevation and atmospheric attenuation. 

Our measurements were composed of pairs of scans 
obtained at different source parallactic angles in each of the years
2011 and  2012. In 2011, the first scan at 17:45 UT on 21 March, lasting
one hour, was followed by a second scan recorded at 20:09 UT the same day, the
source having rotated by approximately $23^{\circ}$ by then. In 2012,
pairs of observations were taken on 21 March and 18 April with corresponding
source parallactic angle rotations of $\approx67^{\circ}$ and $\approx16^{\circ}$,
respectively. All of these pointed observations were collocated with
the HIFI center position. At the same epochs, maps were also produced
that were useful in more accurately describing the maser environment.
A complete summary of the Effelsberg data is provided online in Table \ref{effobssum}. 

\section{Methods/Results\label{sec:Results}}

\subsection{Polarisation Analysis\label{sec:Analysis}}

As explained in \citet{M.Harwit2010}, HIFI does not provide regular
$45^{\circ}$ spacings with which the computation of the Stokes $Q$ and
$U$ parameters is simplest. Instead,
we must rely on the position angles (PA) listed in Table \ref{obssum} for the
vertical polarisation direction, with the horizontal axis at $\mathrm{PA}+90^{\circ}$.
The Stokes $Q$ and $U$ can then be calculated using the analysis
detailed in \citet{M.Harwit2010}, from which the polarisation fraction
and angle are evaluated with

\begin{eqnarray}
p &=& \sqrt{Q^{2}+U^{2}}/I\label{eq:PolFlux}\\
\theta &=& 0.5\arctan(U/Q),\label{eq:polAng}
\end{eqnarray}

\noindent
respectively, with $I$ the total intensity. As explained below, this
analysis has been applied to our entire set of HIFI 621 GHz data, with
the 2011 and 2012 data sets combined to yield a single polarisation
result of sufficient precision. The Effelsberg 22 GHz measurements
were not affected by the same constraints and were considered separately
for the respective 2011 and 2012 data sets. 

\subsection{HIFI data\label{sub:HIFI-data}}

The entire suite of observations listed online in Table \ref{obssum} was
considered, although only the center position of each of the maps
(ObsIDs 1342215920-1 and 1342244411-2), where the maser intensity was 
strongest, was selected for analysis. All data were processed with
version 8 of the \emph{Herschel }Interactive Processing Environment
(HIPE). Following an improvement in the absolute pointing error
(APE) of the \emph{Herschel} pointing products effective 19 Feb 2012
(Observation Day 1011) the reported center of the beam
is accurate to within $0\farcs8-0\farcs9$ ($1\sigma$) for both position switching
and OTF observations. Throughout our observations we found all
pointings to lie within $\simeq0\farcs4$ in Right Ascension and
$\simeq2\farcs4$ in Declination.

As shown in Table \ref{obssum}, the position angles of the telescope
ranged more widely across the observing epochs in 2012 than in
2011. When combined with the 2011 data the three additional datasets 
of 2012 reduced the noise in polarisation intensity ($pI$) from
$\simeq85$ mK to $\simeq17$ mK. Both the Stokes $U$ and $Q$ noise
intensities dropped appreciably as well, to $\simeq32$ and $\simeq14$ mK, respectively. 

\begin{figure}
%\centering
\resizebox{\hsize}{!}{\includegraphics{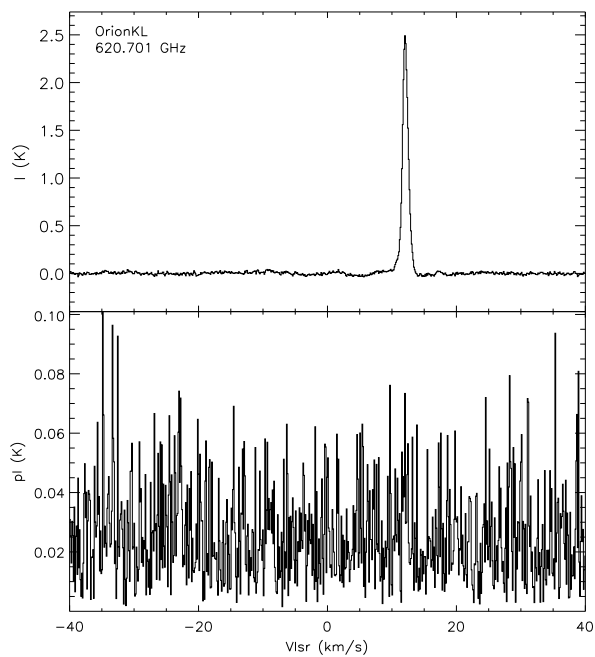}}
\caption{Stokes $I$ (top) and the polarised flux $pI$ (bottom) for the
621 GHz water maser line based on an analysis of all six epochs combined. \label{fig:no_pol}}
\end{figure}

Prior to deriving any polarisation measures, however, a marked change
in the broad spectral component surrounding the narrow maser feature
had to be taken into account. Although each dataset exhibits the broad
component, its strength was clearly higher in the 2012 data, by almost
as much as $50\%$. As explained in Appendix C online, we believe that
this change in the continuum level resulted from small pointing errors
and the proximity of the powerfully emitting Orion ``hot core."  Such
differences, if resulting from pointing errors on an
extended source, can render polarisation studies unreliable. Any claim
about the polarisation of the maser line, therefore, required that it
be separated from the underlying broad feature. To this end, two
Gaussians were simultaneously fitted to the broad component and
removed, to feature the maser line. From epoch to epoch there
remained a substantial variability in the strength of the maser, at
levels similar to those seen in the broad component
itself. Nevertheless, we undertook a polarisation analysis on the
assumption that the maser signals emanate from a single spatially
unresolved  (i.e., point-like) source in the region.  As is
discussed in Section \ref{sub:Pointing-Effects}, this
assumption is consistent with the fact that we found no polarisation
signatures above three standard deviations in the line, at a level of
approximately $2\%$. Figure \ref{fig:no_pol} shows the result of the
analysis after removal of the broad component; the lack of
polarisation in the maser emission is apparent from the absence of a
corresponding signature in polarised flux (pI) in the bottom
panel. The aforementioned change in maser intensity was accompanied by
a shift in the line center velocity of $\simeq0.2$ km s$^{-1}$
across both the horizontal and vertical polarisations from observing
epoch 2 to 3, a significant fraction of the width of the line ($\simeq0.9$ km
s$^{-1}$).

These temporal changes in intensity and velocity are not surprising
since, owing to their location in star forming regions, and in particular at
the forefront of stellar shocks, masers generally have an intensity
that varies strongly on relatively short time scales: it is thought
that such turbulent environs, with their large velocity gradients, can
induce more frequent fluctuations in intensity with changes in the
velocity of the emission line \citep{Stahler2004}.

\subsubsection{Pointing Effects on Maser Polarisation Measurements\label{sub:Pointing-Effects}}

We performed the same polarisation analysis discussed in Section
\ref{sec:Analysis} on a simulated spectrum to quantify the effect
of our pointing uncertainties on measured maser polarisation signals.
To do so the beam profile was approximated by a Gaussian function
matching the FWHM of the HIFI beam at 620.701 GHz
($34\farcs4$), while the source was considered unpolarised and of
no spatial extent (simulating a maser point source). Each simulated
measurement was associated with one of our HIFI observations (see
Table \ref{obssum}), including random pointing errors of about 3\arcsec, such
as might reasonably be expected for HIFI.
The peak intensity of a given measured spectrum would thus decrease
as the pointing position drifted away from the source's location. This
analysis revealed weak polarisation levels averaging $\sim0.7\%$,
consistent with the aforementioned upper limit resulting from our HIFI observations. 

Although we cannot rule out the possibility that the totality of the 621 GHz maser
flux we measure originates from the contributions of several spatially
unresolved sources, our simple simulation seems to imply a lack of
spatial extent for the overall maser emission, as compared to the HIFI
beam. This feature allows for precise polarisation measurements when
the pointing errors are sufficiently small. However, the same is not true for the broad
component of the 621 GHz water line in view of potential source
variations with pointing positions, as discussed in the online Appendix 
\ref{sec:Discussion}. 

\subsection{22 GHz (Effelsberg) Data}

The 22 GHz Effelsberg data, obtained during the same periods (although
not precisely coeval), do not appear to be affected by
the same source of emission responsible for the broad component in
the 621 GHz transition. However, we also observed significant change in
intensity on timescales of one to several months 
(i.e., between respectively 21 March 2012 and 18 April 2012, and  21
March 2011 and 18 April 2012), which imply some intrinsic source
evolution with time. We find that the strongest component at  $7.2$ km
  s$^{-1}$ varies from $\simeq 1.2\times10^{4}$ K to
$\simeq 2.1\times10^{4}$ K (or from 14 kJy to 26 kJy). These
variations are in line with other observations performed at the position
of peak intensity of the 22 GHz maser (located some $\simeq 15\arcsec$
east and $\simeq 36\arcsec$ south from our pointed observations) during
approximately the same time period. More precisely, interferometric
observations of \citet{Matveyenko2012} revealed significant outburst activity between
July 2011 to May 2012 while tracking the evolution of velocity
components at 7.0 km s$^{-1}$ and 7.6 km s$^{-1}$. Previously, in
February 2011, \citet{Hirota2011} recorded a flare reaching an
intensity of 44 kJy using the VLBI Exploration of Radio Astrometry
interferometer. While, unlike our study, these observations were conducted at
very high spatial resolutions ($\sim 1$ mas), single-dish
observations of \citet{Otto2012} at approximately $120\arcsec$
resolution also detected strong flares reaching as high as 80 kJy during an
eight-month period spanning from March to November 2011. 

\begin{figure}
\centering
\resizebox{\hsize}{!}{\includegraphics{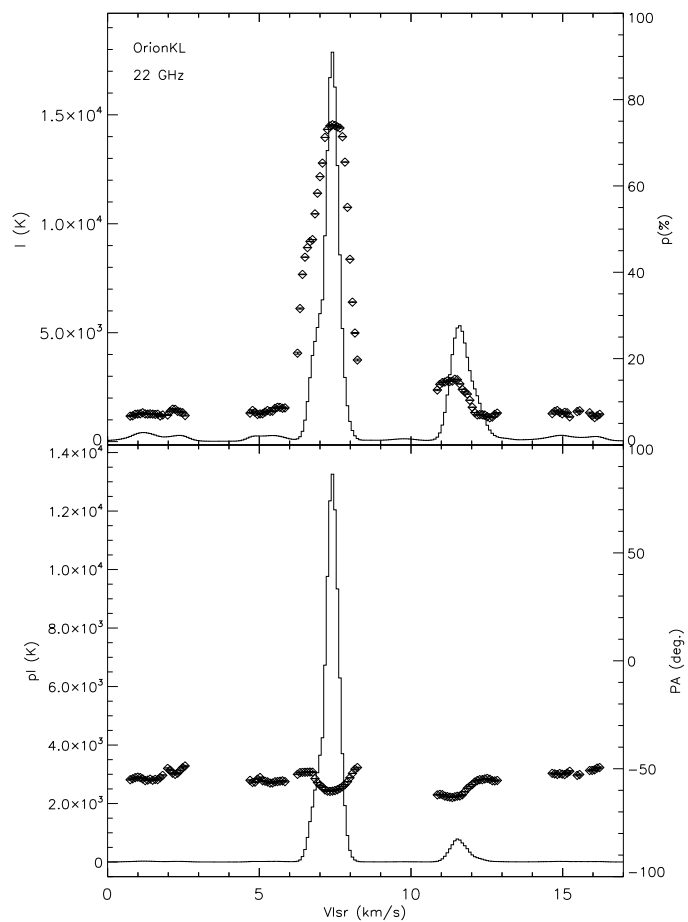}}
\caption{Polarisation properties of the 22 GHz water maser transition as
measured at Effelsberg on 21 March 2011. (\emph{Top)} Overlay
of the percent polarisation (symbols, using the scale on the right)
relative to the total intensity at a velocity resolution of $0.08$ km
s$^{-1}$. (\emph{Bottom)} Polarisation flux and angles (symbols, using
the scale on the right). Only the most prominent features are shown in
each panel.\label{fig:Effpol1}} \end{figure}

We performed a polarization analysis on each of our Effelsberg data
sets and found that the polarization levels did not appreciably change
over the different epochs. Figure \ref{fig:Effpol1} shows the results
of the analysis for the 21 March 2011 data. The top panel displays
the Stokes $I$ spectra, along with several linear polarisation
measures, while the bottom panel shows the polarisation flux and
angles. Most noticeable are the high polarisation levels, of order 
$75\%$ at the center of the $7.2$ km s$^{-1}$ feature. Elsewhere, the
polarisation level is rather constant in the $3-5\%$ range, except for
the second strongest feature at $\sim11.7$ km s$^{-1}$, where levels
exceed $10\%$. While there have been previous detections of
polarisation of comparable levels \citep{Horiuchi2000,Garay1989}, the
aforementioned contemporaneous interferometric observations of
\citet{Matveyenko2012} revealed lower polarization levels of about
$55\%$ at $7.65$ km s$^{-1}$.

\section{Conclusion\label{sec:Conclusion}}

We have reported the astronomical detection of the 621 GHz $J_{K_{-1}K_{1}}=5_{32}-4_{41}$
transition of the ortho H$_{2}$O maser in the star-forming region
Orion KL, which was also recently discussed by Neufeld et al. \citeyearpar{Neufeld2013}.
In observations with the \emph{Herschel/}HIFI instrument 
the maser was found not to be linearly polarised to a 3$\sigma$ upper
limit of $\sim2\%$. 
Low polarisation levels for water masers are not inconsistent with
expectations from the mechanism of Goldreich, Keeley and Kwan \citep{Goldreich1973}, at least in the absence of radiative saturation \citep{Watson2009}.
These results are also approximately in line with the low polarisations
of a few percent found by Harwit et al. \citeyearpar{M.Harwit2010}.

Polarisation measurements of the HIFI data
were complicated by the strengthening of a broad pedestal
component in data from 2011 to 2012. We concluded that this feature is likely
to be the result of pointing artifacts.  Attempts were also made
to settle the change in %polarisation 
maser line strengths between the two observing epochs
but the question remains to be resolved. 
\begin{acknowledgements} 
We thank D. Neufeld et al. (2013) for alerting us to the presence of the 621 GHz transition prior to publication. 
HIFI has been designed and built by a consortium of institutes and
university departments from across Europe, Canada and the United States
under the leadership of SRON Netherlands Institute for Space Research,
Groningen, The Netherlands and with major contributions from Germany,
France, and the US. Consortium members are: Canada: CSA, U. Waterloo;
France: CESR, LAB, LERMA, IRAM; Germany: KOSMA, MPIfR, MPS; Ireland:
NUI Maynooth; Italy: ASI, IFSI-INAF, Osservatorio Astrofisico di Arcetri-
INAF; Netherlands: SRON, TUD; Poland: CAMK, CBK; Spain: Observatorio
Astronomico Nacional (IGN), Centro de Astrobiolog\'{i}a (CSIC-INTA); Sweden:
Chalmers University of Technology - MC2, RSS \& GARD, Onsala Space
Observatory, Swedish National Space Board, Stockholm University -
Stockholm Observatory; Switzerland: ETH Zurich, FHNW; USA: Caltech,
JPL, NHSC. \\
Partly based on observations with the 100-m telescope of the MPIfR
(Max-Planck-Institut f\"{u}r Radioastronomie) at Effelsberg.\\
M.H.'s research is funded through the NSERC Discovery Grant,
Canada Research Chair, and Western's Academic Development Fund
programs. \\
The work of M.O.H. has been supported by NASA through awards of JPL
subcontracts 1393122 and 1463766 to Cornell University. 

\end{acknowledgements}
\bibliographystyle{aa}
\bibliography{/Users/scottjones/Documents/HIFI/paper/research_note/HIFIchpt}

\Online

\begin{appendix}

\section{Summary of Herschel/HIFI and Effelsberg Observations\label{table_summary}}

\begin{table}[ht]
\caption{Summary of HIFI observations conducted in 2011 and 2012.}
\label{obssum}
\centering
\begin{tabular}{lllllll}
\hline\hline
Observation&Observation&Observation&Date&System Temperature&Position Angle\tablefootmark{a}&Mode \\
Number&Day (OD)&ID (ObsID)& &T$_{sys}$ (K) (V/H)&PA ($^{\circ}$) \\
\hline
1V/H & 666 & 1342215918-9 & 2011 Mar 3 & 97/98 & 266.2 & OTF Mapping \\
2V/H & 666 & 1342215920-1 & 2011 Mar 3 & 97/97 & 266.2 & OTF Mapping \\
3V/H & 681 & 1342216838-9 & 2011 Mar 26 & 97/97 & 273.6 & Pointed \\
4V/H & 694 & 1342218904-5 & 2011 Apr 8 & 97/98 & 280.0 & Pointed \\
5V/H & 1017 & 1342239637-8 & 2012 Feb 25 & 68/84 & 258.9 & Pointed \\
6a-V/H & 1066 & 1342244409-10 & 2012 Apr 14 & 68/84 & 284.1 & Pointed \\
6b-V/H & 1066 & 1342244411-2 & 2012 Apr 14 & 68/84 & 284.1 & OTF Mapping \\
\hline
\end{tabular}
\tablefoot{
\tablefoottext{a}{Position of the HIFI vertical polarisation axis relative to north, increasing eastward.}}
\end{table}
 
\begin{table}[ht]
\caption{Summary of Effelsberg observations conducted in 2011 and 2012.\label{effobssum}}
\centering
\begin{tabular}{lllll}
\hline\hline
Observation&Date&System Temperature&Parallactic Angle &	Mode \\
Number	& &T$_{sys}$ (K) (V/H)	&(PA) ($^{\circ}$) \\
\hline
1a-V/H & 2011 Mar 21 & 91/81   & 6.1  & Pointing \\
1b-V/H & 2011 Mar 21 & 115/103 & 28.9 & Pointing	\\
%1c-V/H & 21/03/2011 & 91/82   & -7.3  & Mapping	\\
%1d-V/H & 21/03/2011 & 118/104 & -29.6 & Mapping	\\
2a-V/H & 2012 Mar 21 & 108/95 &  -30.3 & Pointing \\
2b-V/H & 2012 Mar 21 & 134/117 & 37.0 & Pointing \\
%2c-V/H & 21/03/2012 & 91/79 &  29.7 & Mapping	\\
%2d-V/H & 21/03/2012 & 86/74  & -8.5  & Mapping	\\
3a-V/H & 2012 Apr 18 & 94/81 & 19.2 & Pointing \\
3b-V/H & 2012 Apr 18 & 136/118 & 34.9 & Pointing \\
%3c-V/H & 18/04/2012 & 102/89 & -20.0 & Mapping \\
\hline
\end{tabular}
\end{table}

\clearpage

\section{Herschel/HIFI Maps of 621 GHz Water Maser\label{HIFImaps}}

\begin{figure}[ht]
%\centering
\resizebox{\hsize}{!}{\includegraphics{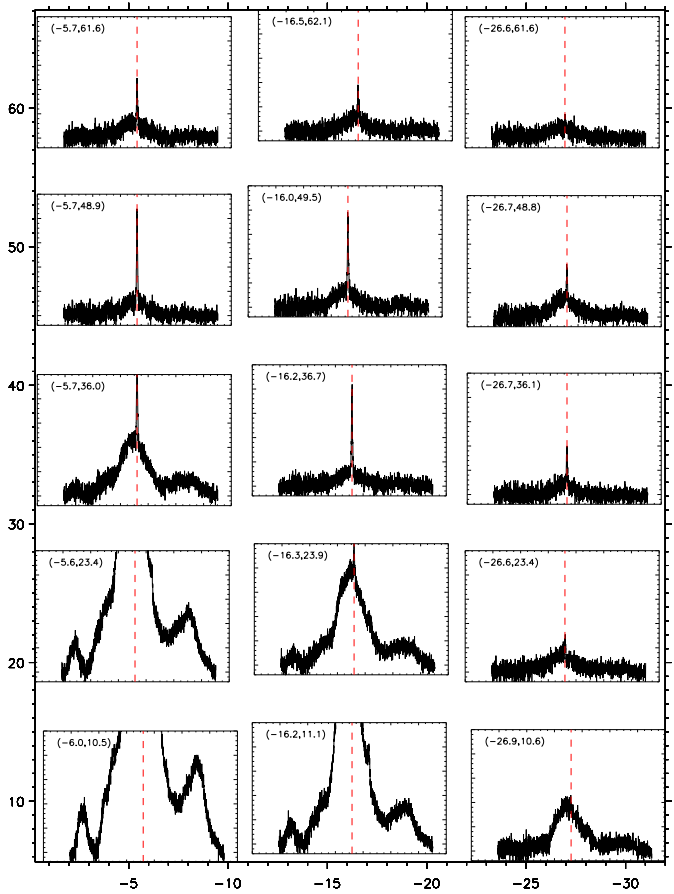}}
\caption{Map of 621 GHz emission corresponding to dataset 2-H in Table
\ref{obssum}. The offsets from (R.A.{[}J2000{]}$=05^{\mbox{h}}35^{\mbox{m}}14^{\mbox{s}}.3$,
decl.{[}J2000{]}$=-05^{\circ}22\arcmin33\farcs7$) are given in the
upper left of each panel in seconds of arc. The horizontal and
vertical axes bordering the entire set of 15 panels correspond roughly
to these respective offsets for the positions at which the spectra
shown in the individual panels were observed. The fine vertical scales
on the individual panels, run from $[-0.2 \mbox{ to } 2.5]$ degrees Kelvin; the
width of the individual panels cover a $V_{lsr}$ range of $[-60
\mbox{ to } 80]$ km s$^{-1}$
roughly centered on the vertical red line marking $V_{lsr} = 12 \mbox{ km
s}^{-1}$, the velocity of the narrow 621 GHz maser feature.\label{fig:Map_first}}
\end{figure}

\begin{figure}
%\centering
%\resizebox{\hsize}{!}
\begin{flushright}
\includegraphics[scale=0.45]{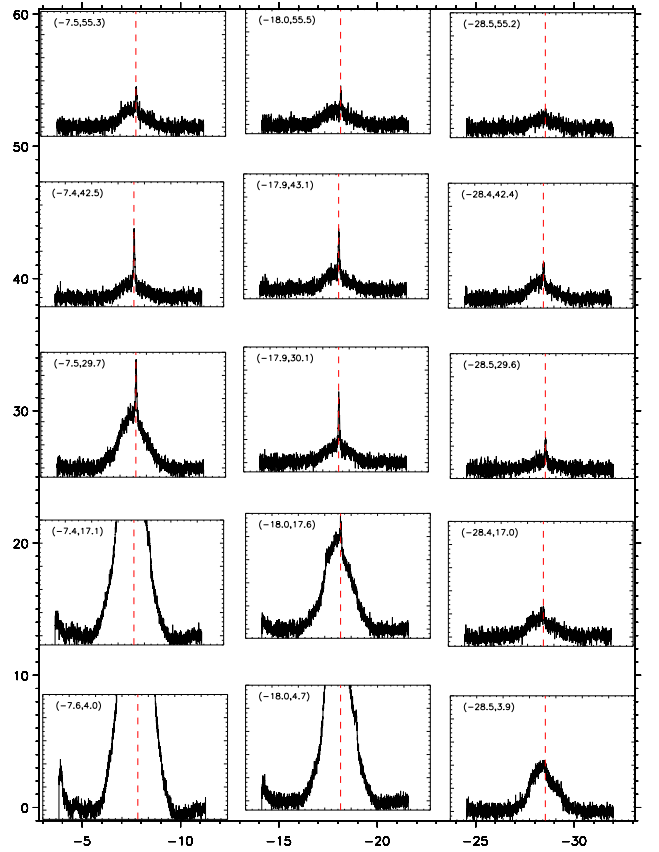}
\end{flushright}
\caption{Same as \ref{fig:Map_first} but for dataset 6b-H in Table
\ref{obssum}. \label{fig:Map_last}}
\end{figure}

\section{Discussion on Pointing Errors\label{sec:Discussion}}

%\subsection{Temporal Variations}

The source for the variability of the broad pedestal component in the 621
GHz spectral line requires explanation.
More precisely, we need to determine whether some time-varying
physical mechanism and/or pointing errors are responsible for the change that occurs in 
the pedestal between 2011 to 2012.

As the first and last set of observations involved a small map, we
examined two possibilities. First, whether the increase in emission
from the broad component is common to all areas of the map, and
second, if not, whether there is a noticeable offset in the spatial
positioning. Figures \ref{fig:Map_first} and \ref{fig:Map_last}
illustrate the horizontally polarised versions of the two small maps
taken during the first and last epochs of our observations. In Table
\ref{obssum}, this corresponds to observation numbers 2-H and 6b-H,
respectively.  
From 2011 and 2012, 
the intensity of the maser line itself
seems to systematically decrease between the maps of Figures \ref{fig:Map_first}
and \ref{fig:Map_last}.  At the same time, in many panels toward the
map center, the aforementioned broad component appears to get stronger.
This trend differs at points farther from the center, where there
is more of a decrease, especially in the line wings of the bottom
left panels. This latter point seems to indicate systematic changes
in the source. However, one possible explanation for the increase
in the broad component may be due to the proximity of the Orion
``hot core'' (HC) whose molecular line emissions were studied extensively
by \citet{Beuther2005}. The HC is located at (R.A.{[}J2000{]}$=05^{\mbox{h}}35^{\mbox{m}}14^{\mbox{s}}.50$,
decl.{[}J2000{]}$=-05^{\circ}22\arcmin30\farcs45$) only about 6 arcseconds 
removed from Orion KL. It is displaced
from our pointing direction by only $20\arcsec$ in Right Ascension
and $30\arcsec$ in Declination, and thus lies at an offset of only
36 seconds of arc from our prime pointing direction. This small offset,
barely exceeding the FWHM of our beam at 621 GHz, implies that
even a relatively small error in pointing could effect a large apparent
variability, given that the HC would lie on the flank of our beam
profile.

In Section \ref{sub:HIFI-data}, we stated that the adjustment
in the absolute pointing error (APE) improved the pointing accuracy of
HIFI but retained some of its uncertainty. This is especially true of the mapped observations,
which have additional pointing uncertainties due to relative offsetting
and jitter, that may total $1\farcs5$-$2\farcs5$ for OTF maps. Both
mapping and pointed modes can be further misdirected
from their intended pointing during telescope switching from OFF to
ON source. Indeed the APE was verified during \emph{Herschel }photometric
operation and was never proven to exactly match while in spectroscopic
observations. In general, HIFI has shown itself able to resolve source
structure to better than $1\arcsec$ at its high-frequency end.
Nevertheless pointing errors of order $3\arcsec$ have occasionally
been observed on \emph{Herschel}.

As we are combining data across several Observation
Days, it is easy to conceive of pointing errors of this order.
Together with the large source gradient shown
in Figures \ref{fig:Map_first} and \ref{fig:Map_last} and the presence
of the HC in the vicinity of our source, the effect of pointing errors
must be considered a likely source of variability 
in the broad spectral component. Quantitatively, if one is to look
more closely at, for example, Figure \ref{fig:Map_last}, there is
a drastic, $\sim1.5$ K increase in the broad component as one moves
$12\farcs5$ from the center position to that directly beneath. Following
the discussion from above, a reasonable pointing error of $3\arcsec$,
or 23\% of the map step size, would thus correspond to an increase
of $\sim0.35$ K in the broad pedestal. An offset of this magnitude
from the center position would therefore result in the amplitude of
the broad component rising to $\sim0.85$ K, or $\sim1.7$ times the
intensity.

\end{appendix}

\end{document}